%% file: main.tex
\documentclass{article}

\usepackage[a4paper, margin=0.8in]{geometry}
\usepackage[utf8]{inputenc}
\usepackage[dvipsnames,table,xcdraw]{xcolor}
\usepackage[affil-it]{authblk}
\usepackage[misc]{ifsym}
\usepackage{
amsmath,
amssymb,
amsthm,
appendix,
arydshln,
braket,
caption,
csquotes,
enumitem,
fdsymbol,
fontawesome,
forest,
graphicx, 
hyperref,
kantlipsum,
listings,
longtable,
mathtools,
multicol,
multirow,
qcircuit,
rotating,
sectsty,
subcaption,
subfiles,
tcolorbox,
tikz,
ulem,
url,
verbatim,
xcolor,
ulem}

\tcbuselibrary{raster, skins, breakable, listings}
\def\titlefont{\color{RoyalPurple}}
\sectionfont{\color{RoyalPurple}}
\subsectionfont{\color{RoyalPurple}}
\subsubsectionfont{\color{RoyalPurple}}

\let\OLDthebibliography\thebibliography
\renewcommand\thebibliography[1]{
  \OLDthebibliography{#1}
  \setlength{\parskip}{0pt}
}

\title{\titlefont \textbf{\LARGE The computational inevitability of life}\\\textbf{\large self-replication under resource-bounded nested algorithmic probability}}

\author[1,2]{Aritra Sarkar}

\affil[1]{Department of Quantum \& Computer Engineering, Delft University of Technology, The Netherlands}
\affil[2]{Quantum Intelligence Alliance, Kolkata, India}

\affil[ \Letter ]{aritra.sarkar@quantum-intelligence.net}

\date{}

\begin{document}


\maketitle

\input{content}



\section*{Acknowledgements}

The authors thank Marcus Hutter, Zaid Al-Ars, David DiVincenzo, and Harry Buhrman for insightful discussions regarding the formalization of bounded algorithmic probability in an earlier draft~\cite{sarkar2020quines} (irony unmissed, this article cites itself). 

The authors used ChatGPT and Grammarly as AI-assisted editing tools to improve grammar, clarity, and readability during manuscript preparation. 
The manuscript has been carefully reviewed by the authors, and they take full responsibility for the final content.

\bibliographystyle{unsrt}
\bibliography{ref.bib}


\end{document}

%% file: content.tex
\begin{abstract}
Recent computational experiments have demonstrated the spontaneous emergence of self-replicating programs across universal automata, artificial chemistries, and self-modifying code systems. Remarkably, these results arise without explicit fitness functions, reward shaping, or predefined objectives, indicating a gap in our formal understanding of the underlying computational process.

In this work, we argue that self-replication is computationally inevitable under resource-bounded automata. 
Building on algorithmic information theory, we show that when universal inductive bias is applied under finite constraints of time, memory, and description length, programs that construct descriptions of themselves, i.e., quines, emerge as stable fixed points of nested algorithmic probability. 
We formalize this argument and demonstrate that self-replicating programs act as attractors in program space, independent of external optimization criteria. 
Thus, resource bounds transform universal induction into a competitive ecological process over programs, in which self-constructing programs dominate by stabilizing their own measure under resampling.

We reinterpret recent results from computational life experiments and self-improving artificial agents as empirical realizations of this theoretical principle.
More broadly, we propose that life is the simplest persistent structure available to constrained computation.
A living system remembers itself because doing so is algorithmically and thermodynamically unavoidable.
\end{abstract}


\section{Introduction}
\label{sec_1}


Recent computational experiments have demonstrated that life-like behavior, in particular self-replication, repeatedly emerges in abstract computational systems. Notably, \cite{alakuijala2024computational} demonstrates that self-replicating programs arise spontaneously in minimal instruction-set languages initialized with random programs, without explicit fitness functions, reward shaping, or predefined objectives. 
These systems exhibit sharp phase transitions characterized by entropy collapse and the dominance of replicators, with replication emerging prior to any form of selection or optimization. 
Similar phenomena have been observed in other artificial life and self-modifying program ecosystems~\cite{ray1996approach,adami1998introduction,langton1986studying}, suggesting that replication is a robust and substrate-independent feature of computation rather than an artifact of biological chemistry or engineered incentives.

These observations motivate the central research question of this work: \textit{Why does self-replication keep arising across unrelated computational systems?}
The recurrence of replication across systems that differ radically in architecture, semantics, and dynamics suggests a deeper organizing principle rooted in the structure of computation itself.
Existing theoretical frameworks do not fully account for this phenomenon.
For instance:
(i) Darwinian evolution presupposes the existence of replicators and therefore cannot explain their origin; 
(ii) optimization and reinforcement learning assume an externally specified cost function and rewards, which are absent in the systems under consideration; 
(iii) thermodynamic approaches, while successful in describing entropy production and dissipation, generally lack symbolic and algorithmic structure, and, 
(iv) chemical origin-of-life models, though insightful, are substrate-specific and do not generalize to abstract automata. 
Consequently, none of these approaches explains why replication appears so ubiquitously in computational universes.

In this work, we argue that self-replication is a computational inevitability under resource-bounded automata. 
Building on algorithmic information theory~(AIT), we show that when universal inductive bias is applied under finite constraints of time, memory, and description length, programs that construct descriptions of themselves, i.e., quines or constructors, emerge as stable fixed points of our proposed metric: resource-bounded nested algorithmic probability. 
Under such constraints, universal induction induces a competitive ecological process over programs, in which self-constructing programs dominate by stabilizing their own measure under repeated resampling. 
This provides a substrate-independent explanation for the emergence of replication that does not rely on optimization, selection, or externally imposed goals.

The remainder of the paper is organized as follows. 
Section~\ref{sec_2} provides the necessary background and reviews related work. 
In Section~\ref{sec_3}, we introduce the proposed AIT metric of resource-bounded algorithmic probability and its nested variant. Thereafter, we conduct empirical studies demonstrating the dominance of self-replicating programs as fixed points under conditions of resource constraints and nesting. 
Section~\ref{sec_4} discusses the implications of these results for biology, artificial general intelligence, and fundamental physics.

\section{Background and related works}
\label{sec_2}


This section surveys prior theoretical and empirical studies on self-replication in abstract computational systems, spanning constructor theory, artificial life, cellular automata, and recent large-scale computational experiments. 
We emphasize their shared structural features and position them as converging evidence for a deeper computational principle underlying the emergence of replication as addressed by this article.

\subsection{Constructors, quines, and artificial life}

The earliest formal treatment of self-replication is due to John von Neumann, who introduced the notion of a self-reproducing automaton capable of constructing a copy of itself given a symbolic description~\cite{neumann1966theory}. 
Von Neumann's construction relies on the separation between a constructor and a description, anticipating the logical structure of biological replication in which DNA functions as a symbolic control tape. 
Importantly, this framework does not assume evolutionary optimization; replication is defined as a constructive fixed point in a universal automaton.

In modern theoretical computer science, von Neumann constructors can be understood as realizations of universal Turing machines~(UTM) equipped with a description-processing mechanism. 
Kleene's recursion theorem formalizes this idea by guaranteeing the existence of programs that can refer to and reproduce their own descriptions~\cite{kleene1936general}. 
In programming languages, these fixed points manifest as quines: programs that output their own source code. 
Quines represent the software analogue of von Neumann constructors on a substrate of universal computation.

Artificial life~(ALife) research has extensively explored the emergence and dynamics of self-replicators in computational environments. 
Early ALife systems such as Langton's cellular automata~\cite{langton1986studying} and Ray's Tierra~\cite{ray1996approach} demonstrated that self-replicating programs can arise and evolve under minimal assumptions. 
Subsequent work emphasized the role of information processing, program size, and mutational robustness in sustaining replication~\cite{adami1998introduction}. 
However, most ALife systems still rely on explicit selection pressures, fitness proxies, or survival criteria, leaving open the question of why replication appears even in their absence.

\subsection{Spontaneous replication in minimal computational substrates}

Recent work by Alakuijala et al. provides empirical evidence that self-replication emerges robustly in minimal computational systems~\cite{alakuijala2024computational}. 
In these experiments, populations of random programs were executed in a simple register-based language with bounded memory and instruction sets. 
No explicit fitness function, reward signal, or selection objective is imposed. 
Well-formed self-replicating programs were found to arise spontaneously and rapidly dominate the population.

A key observation in this work is the presence of a sharp phase transition in program-space dynamics. 
Initially, program executions exhibit high entropy and transient behavior, with most programs terminating or producing unstructured outputs. 
Beyond a critical threshold in population size or execution depth, the system undergoes an entropy collapse characterized by the sudden appearance and persistence of replicators. 
Replication precedes any form of optimization or competition, indicating that selection operates only after constructors are already present.

It is important to note that the observed replicators are not hand-designed quines but arise naturally from random initialization, suggesting that self-replication is a generic attractor in bounded computational systems. 
The authors further show that replicators compress the reachable state space by stabilizing their own descriptions, thereby biasing subsequent executions toward low-entropy regions of program space. 
This behavior closely mirrors earlier theoretical predictions concerning measure concentration under algorithmic probability~\cite{solomonoff1964formal,li2008introduction}.
The empirical instantiation in \cite{alakuijala2024computational} motivates our exploration for an underlying theoretical principle.

\subsection{Metabiology and assembly theory}

An explicit attempt to recast biological evolution in computational terms is found in Chaitin's program of metabiology~\cite{chaitin2013proving}, which models evolution as an algorithmic process driven by program-size complexity rather than physical fitness alone. 
In this framework, organisms are represented as programs, mutations correspond to program modifications, and evolutionary progress is measured in terms of increasing algorithmic complexity. 
It emphasizes that evolutionary novelty is fundamentally constrained by computability and algorithmic randomness, framing biology as a special case of open-ended computation.
While metabiology provides a principled algorithmic abstraction of evolution, it presupposes the existence of self-replicating entities and focuses primarily on post-replicative dynamics. 
Thus, it does not directly address why replication itself should emerge in the first place. 
Nevertheless, its core insight of biological processes being governed by computational limits imposed by AIT aligns closely with our present work.

More recently, assembly theory~\cite{sharma2023assembly} proposes a quantitative measure of complexity based on the minimal number of steps required to assemble an object from simpler components. 
It aims to provide a substrate-independent criterion for distinguishing living systems from non-living ones, emphasizing the causal and historical structure of construction, termed assembly pathways, rather than statistical complexity. 
Although assembly theory is formulated operationally rather than algorithmically, it admits a natural reinterpretation in AIT terms. 
Assembly pathways correspond to constructive descriptions, and assembly indices act as upper bounds on Kolmogorov complexity under constrained generative grammars.
It has been argued~\cite{abrahao2024assembly} that measures of biological and physical complexity ultimately reduce to algorithmic probability rather than purely statistical or combinatorial metrics.
The present work extends this perspective by showing that, under resource-bounded nested algorithmic probability, such constructive fixed points are not exceptional but inevitable. 
In this sense, assembly, replication, and life emerge as different manifestations of the same underlying algorithmic principle: persistence under constrained computation.

\subsection{Self-modification and continual novelty}

Darwin--G\"odel machines~(DGMs), is a recently introduced class of self-modifying computational agents~\cite{schmidhuber2007godel} that improve their own code through iterative proposal, evaluation, and incorporation of modifications~\cite{zhang2025darwin}. 
DGMs are not guided by a fixed objective function or externally specified reward signal. 
Instead, they operate through a population-based evolutionary process over programs, where candidate self-modifications are generated, tested, and retained if they lead to improved future performance. 
The system allows unrestricted self-referential modification, enabling agents to rewrite the very mechanisms by which they evaluate and transform themselves.
A key insight from DGMs is that self-referential improvement emerges naturally once programs are permitted to act on their own descriptions. 
From this work's perspective, DGMs operate in a regime where replication and self-description are already present, and the primary challenge is avoiding stagnation at trivial fixed points. 
The success of DGMs, therefore, validates the present thesis: self-referential constructors are not engineered artifacts but natural attractors of computational systems once resource-bounded execution and description manipulation are allowed.

Another complementary work study, Digital Red Queen~(DRQ) dynamics, demonstrates that sustained novelty and adaptation can arise in computational ecosystems without explicit objectives or external reward shaping~\cite{kumar2026digital,kumar2025automating}. 
In these systems, agents co-evolve through relative performance comparisons, with progress defined only with respect to other agents. 
This setup produces continual arms-race dynamics reminiscent of biological Red Queen evolution, where maintaining competence requires ongoing adaptation even in a stationary environment.
DRQ dynamics highlight that without relative competition or environmental feedback, systems tend to collapse into static equilibria. 
From the standpoint of algorithmic information theory, such a collapse corresponds to domination by minimal fixed points in program space. 
The relevance to the present work is twofold. 
First, DRQ dynamics explain how open-ended evolution can persist after the emergence of self-replicating constructors. 
Second, they reinforce the claim that replication itself is prior to selection and optimization.

\subsection{Self-replication in cellular automata}

Cellular automata (CA) are a canonical model for studying the emergence of complex behavior from simple local rules. 
Early work by von Neumann~\cite{neumann1966theory} and Langton~\cite{langton1986studying} established that self-replication is possible in CA given sufficient rule complexity and symbolic structure. 
More recently, \cite{cotler2025self} provides a rigorous and unifying treatment of self-replication in cellular automata by explicitly linking it to two kinds of Turing universality for CA: local, i.e., an automaton's ability to simulate any Turing machine, and global, i.e., the ability to simulate any same-dimension CA. 
This distinction becomes fundamental in addressing self-replication: global universality implies self-replication; local universality doesn't.
Importantly, their construction does not rely on evolutionary dynamics, selection pressure, or optimization, reinforcing the idea that replication precedes Darwinian processes.
From the perspective of the present work, these results provide complementary evidence for the inevitability of replication in abstract computational systems. 

\section{Formalization and empirical results}
\label{sec_3}


In this section, we formalize the central claim of this work: that self-replication emerges inevitably under universal induction when computation is subject to finite resource constraints. 
By extending algorithmic probability to explicitly incorporate bounds on time, memory, and description length, we show how universal induction induces an ecological competition over programs, in which self-constructing programs arise as stable fixed points.

\subsection{Resource-bounded algorithmic probability}

Algorithmic information theory~\cite{li2008introduction} considers that the length of data is a false indicator of its complexity.
Instead, the program's length is used to calculate the likelihood of the data.
This is dichotomous in the von Neumann stored-program perspective, as the reasoning that applies to data can be extended to programs as well, treating them as a form of data.
AIT does not consider how likely it is that short programs are generated by a higher-level (meta-)program.
The formalism we develop addresses this dichotomy.

We consider a fixed-length model (i.e., a universal linear bounded automata), where the data, the program, and all higher-level programs have the same length, and are input programs or outputs of the same automata.
This automaton is denoted by $\mathcal{A}$.
The length is a resource limitation from any realistic automata implementation.
For naturally occurring computing hardware such as DNA or the brain, it is imperative to have a fixed resource for storing the program, e.g., the number of base pairs or neurons.
We explore the properties of the final distribution of the data string, provided we consider multiple levels of meta-programs.
The distribution converges to self-replicating programs, or quines, which survive across generations of the program-data hierarchy.

The universal Solomonoff algorithmic probability~\cite{solomonoff1964formal} of a program $p$ on a (prefix-free) universal Turing machine (UTM) $U$ for an output $x$ is proportional to the sum of the inverse of the description lengths of all programs that generate the output.
$$\xi_U(x)=\sum_{p:U(p)\rightarrow x} 2^{-l(p)}$$
This naturally formalizes Occam's razor (law of parsimony) and Epicurus's principle of multiple explanations by assigning larger prior credence to theories that require a shorter algorithmic description while considering all valid hypotheses.

Consider a Turing machine $T$ with $n$ symbols and $m$ states, which can be enumerated by running a finite number of programs $p$.
The algorithmic probability of a string $x$ can be approximated as:
$$\xi(x) \approx D_{n,m}(x)=\dfrac{|T\in(n,m): T(p) \text{ halts with output } x|}{|T\in(n,m): T(p) \text{ halts}|}$$
i.e., by counting how many programs produce the given output divided by the total number of programs that halt.
The $|.|$ is used to denote set cardinality throughout this section.
This approximation is called the Coding Theorem Method (CTM)~\cite{levin1974laws}.


We are interested in the distribution of the computing output generated by a set of fixed-size programs.
We do not consider a special halt state, thus allowing us to explore the complete state space of programs~\cite{sarkar2020quantum} (i.e., the powerset of the full language of fixed length).
This can encode programs that demonstrate a halting behavior (i.e., the tape output does not change after a certain time) by encoding the halting TM state as states of another TM that loop on themselves, move the tape head arbitrarily, and write back the read character.

Let the description number of $n$ symbols and $m$ states be encoded as binary strings of length $l$.
Thus, all $2^l$ possible programs have, and when run for $t$ time steps (of course, preferably larger than the Busy Beaver number), produce the following approximation of algorithmic probability:
$$D^{tl}_{n,m}(x)=\dfrac{|\mathcal{A}^t(p) \rightarrow x|}{2^l}$$

We reach the same result by plugging in the constant size of programs in the original equation of $\xi(x)$. Note that, in the fixed length and time case, the automata is always prefix-free but no longer guaranteed to be universal.
$$\xi(x)_\mathcal{A} \approx D^{tl}_{n,m}(x)=\sum_{p:\mathcal{A}(p)\rightarrow x} 2^{-l}=2^{-l}|\mathcal{A}(p)\rightarrow x|$$

\subsection{Nested algorithmic probability}

Let's denote the resource-bounded algorithmic probability derived in the last section as:
$$\xi^{01}(x^0)=2^{-l}|\mathcal{A}(x^1)\rightarrow x^0|$$
$\xi^{01}$ denotes the algorithmic probability of the set of output strings $x^0$ and is based on considering a uniform distribution of the set of programs $x^1$.
We drop the subscript $\mathcal{A}$ for brevity, but all results should be interpreted as having a dependence on the automata.
This scalable notation considers fixed-length programs and output data, no inputs, and a specific automata $\mathcal{A}$.
Thus, the set cardinalities are $|X^1|=|X^0|=2^l$.
Individual lengths of the strings, $x^1 \in X^1$ and $x^0 \in X^0$ are $l(x^1)=l(x^0)=l$.

\begin{figure}[ht]
    \centering
    \includegraphics[clip, trim=5cm 0cm 0cm 7cm,width=0.7\columnwidth]{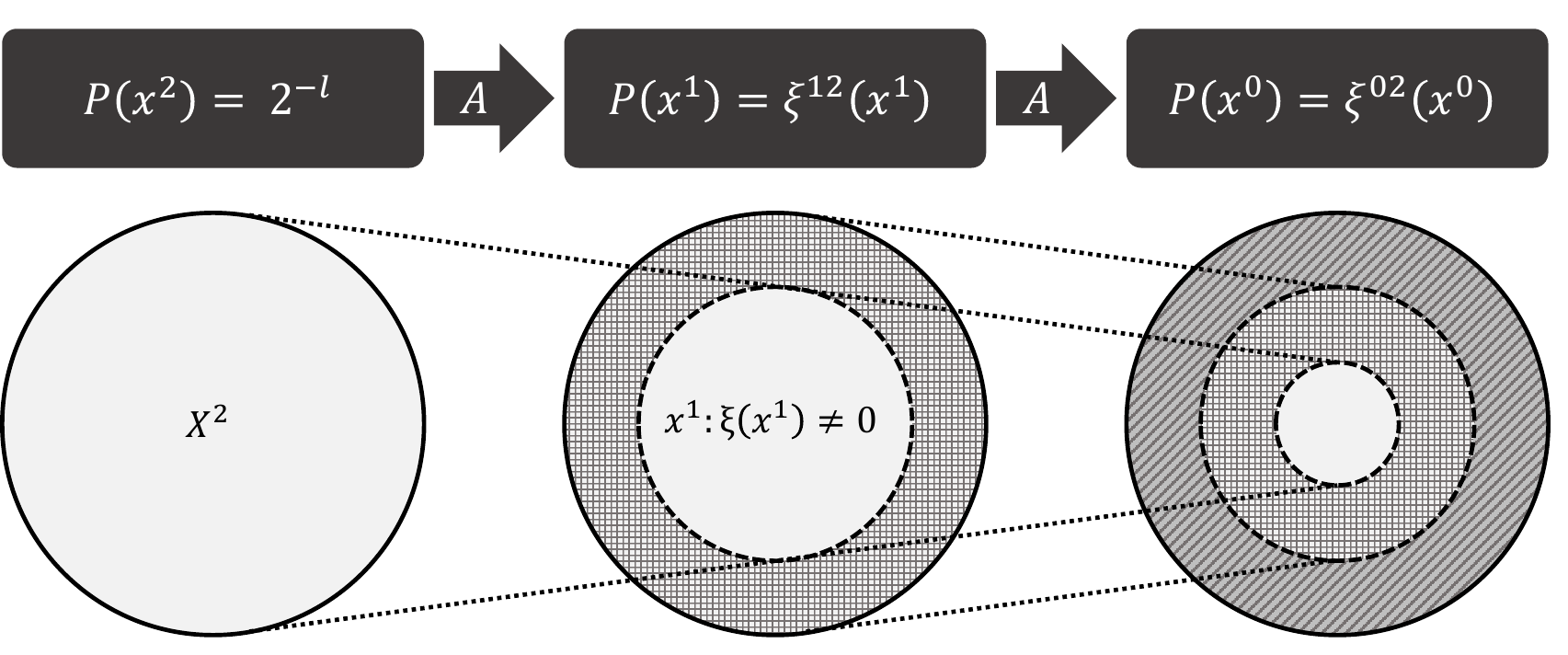}
    \caption{The space of programs typically maps to a smaller set of output strings. The algorithm probability $P(x^0)$ of the output strings $x^0$ is based on a uniform probability $2^{-l}$ of the programs in the highest meta-level (left circle).}
    \label{fig:level01}
\end{figure}

In general, this is a many-to-one mapping; thus, not all strings in $X^0$ are generated by running programs of the same size.
The strings that are algorithmically random are not part of the set $X^0$ and are shown as the striped annulus in Figure~\ref{fig:level01}.

The set of algorithmic probabilities $\xi^{01}$ for all $X$ is the resource-bounded approximation of Solomonoff's universal (prior probability) distribution.
The core motivation of algorithmic information theory to define the universal distribution from a uniform distribution is based on the idea that simpler, shorter theories are more probable.
Our fixed-length program (model/hypothesis) formulation does not seem to allow shorter programs to have more weight.
Yet, there will be more programs to generate simple data; e.g., $0101010101010101$ can be generated by looping $01$ 8 times, $0101$ 4 times, or $01010101$ twice.
Whichever of these 3 programs is most efficient would print the output and reach a halting state early.
What matters is the frequency of programs (as in the CTM), not at what stage in the computation of our $t$ steps it reached a stable attractor state.
So even within the same length programs, we expect a non-uniform distribution as we do for the data.

Now we can pose the question: `Why should all programs be considered equiprobably?'
Considering a higher hierarchy of meta-programming, there is some physical process that generates that program on the automata's program tape.
Normally, in the standard definition of algorithmic probability, it is considered an unbiased coin flip or the infinite-programming-monkey theorem.
Since the universal distribution is not known a priori, there is no other preference than a uniform distribution to start with.

Thus, the generalized form of the previous equation would be:

$$\xi^{02}(x^0)=2^{-l} \sum_{x^1: \mathcal{A}(x^1)\rightarrow x^0} \Big[ \xi^{12}(x^1). 2^l \Big] = \sum_{x^1: \mathcal{A}(x^1)\rightarrow x^0} \xi^{12}(x^1)$$

If $\xi^{12}(x^1) = 2^{-l}$ for all $x^1 \in X^1$, the definition would converge to the previous case, where the summation is simply counting the number of programs with the property $\mathcal{A}(x^1)\rightarrow x^0$.

However, we can feed back the universal distribution into the programs and understand its effect on the data.
Thus introducing a hierarchy of automata levels, where the output data of one automata is the input program for the next level.
Weighting the contribution of each program based on the probability that they themselves physically occur in the program part of the tape.

Due to the invariance theorem, we can assume that the automata for the two levels are not same, with only constant overheads of translating (cross-compiling) the program of one machine to another, however, for our case, we use the same automata $\mathcal{A}$, i.e. $\xi_{\mathcal{A}_0}^{01}(x^0)=\xi_{\mathcal{A}}^{01}(x^0)$ and $\xi_{\mathcal{A}_1}^{12}(x^1)=\xi_{\mathcal{A}}^{12}(x^1)$.

We used the superscript $02$ to distinguish it from the earlier standard definition (with the superscript $01$). 
This denotes that the final output is at level $0$, while there are $2$ levels of execution on the automata $\mathcal{A}$ that lead to this distribution, or $2$ meta-levels.
Note that now, $\xi^{23}(x^2) = 2^{-l}$; thus, while the programs (or level 1 meta-programs) $x^1 \in X^1$ are no longer equiprobable, the level 2 meta-programs $x^2 \in X^2$ are drawn uniformly at random.

Continuing this analogy, adding another meta-level would be of the form:

$$\xi^{03}(x^0) = \sum_{x^1: \mathcal{A}(x^1)\rightarrow x^0} \xi^{13}(x^1) = \sum_{x^1: \mathcal{A}(x^1)\rightarrow x^0} \Bigg[ \sum_{x^2: \mathcal{A}(x^2)\rightarrow x^1} \xi^{23}(x^2) \Bigg]$$

Indeed, it is possible to extend this argument to an arbitrary many levels $w$, with no particular reason to choose $\xi^{01}$ over $\xi^{0w}$ for the expected distribution of the data occurring physically.
This generalized recursive algorithmic probability is defined as:

$$\xi^{ab}(x^a) = \sum_{x^{a+1}: \mathcal{A}(x^{a+1})\rightarrow x^a} \xi^{(a+1)b}(x^{a+1})$$

where, $\xi^{(w-1)w}(x^{w-1}) = 2^{-l}$, and the base case is a uniform distribution.

Let the number of strings $N^{0w} \le 2^l$ with non-zero probability for a particular meta-level $w$ be defined as:
$$N^{0w} = |\xi^{0w}(x^0) > 0| = \sum_{x^0 \in X^0} \lceil \xi^{0w}(x^0) \rceil $$

It can be easily seen that, since the programs are deterministic (i.e., have only 1 output), the program-data input-output map of the automata on the space of $2^l$ bit strings is a non-injective, non-surjective function in the general case. 
Thus, for $r_1 > r_2$, $N^{0r_1} \le N^{0r_2}$.
We are interested in the properties as $w$ grows.

\begin{itemize}[nolistsep,noitemsep]
    \item What are the properties of the strings that `survive' over these generations (meta-levels) by being stable under resampling?
    \item Since each added hierarchy reduces the set size, at some threshold value of $r_1 > r_2 > w_t$ the inequality will become an equality $N^{0r_1} = N^{0r_2}$. Is $N^{0r_1} = N^{0r_2} \ne 0$ in that threshold?
\end{itemize}

This formalism of nested resource-bounded universal induction can be viewed as a competitive ecological process over programs~\cite{kumar2026digital}, with programs as genotypes, execution as phenotypes, and output programs as offspring.

\subsection{The algorithmic fitness of quines}

A quine is a program that takes no input and produces a copy of its own source code as its output.
It may not have other useful outputs.
In computability theory, such self-replicating (self-reproducing or self-copying) programs are fixed points of an execution environment, as functions that transform programs into their outputs.
The quine concept can be extended to multiple meta-levels, called ouroboros programs or quine-relays.
Quines are also a limiting case of algorithmic randomness, as their length is the same as their output.
In the case of Turing machine automata, this corresponds to printing the description on the tape, which can be executed as a program by another Turing machine.
Note that the rest of the Turing machine mechanism, such as the tape head and movement, is akin to the underlying cellular automata rules that automatically apply to the new cells where the replicated machine manifests.
The very design of the 3 parts of a constructor suggests that it cannot be algorithmically random, as it should be possible to compress parts of its description.
We are interested in the complexity and probability of constructors, which form a subset of all possible configurations an automaton can possess.

\begin{figure}[ht]
    \centering
    \includegraphics[clip, trim=13cm 0cm 0cm 5cm,,width=0.7\columnwidth]{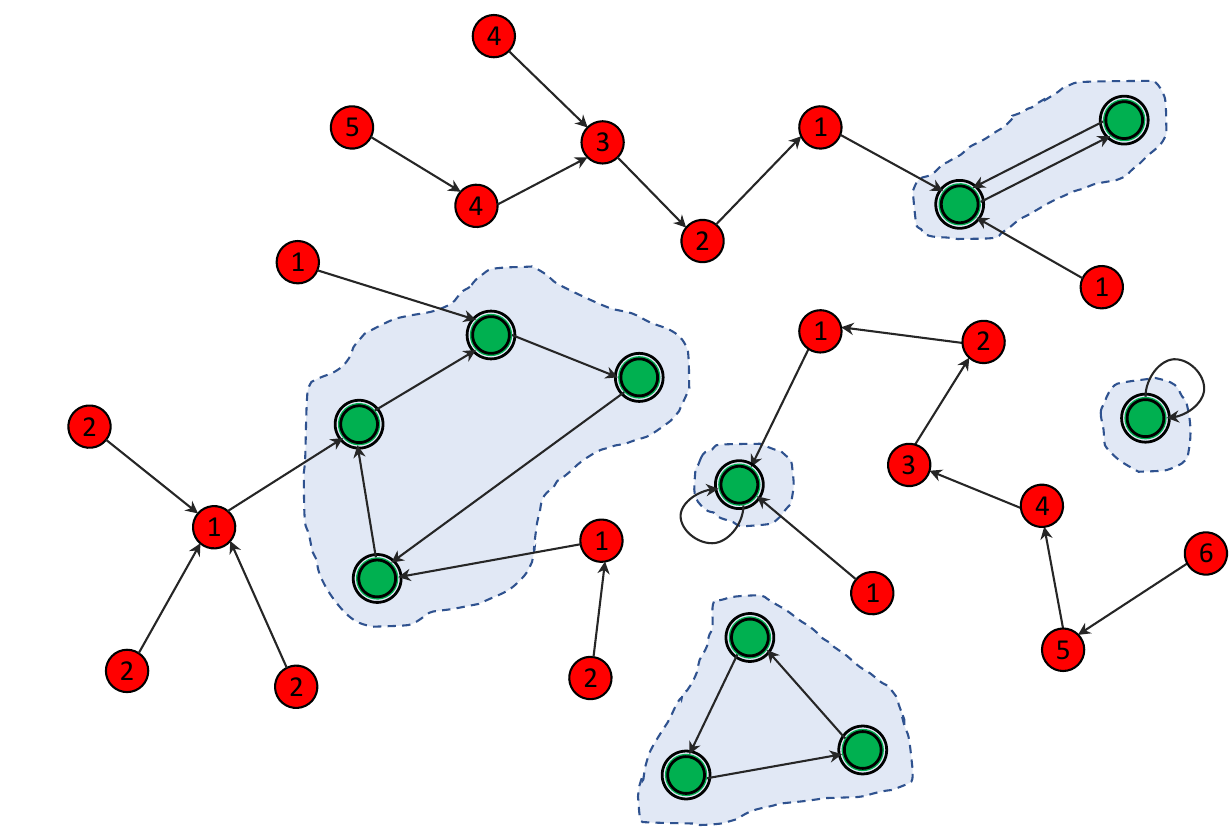}
    \caption{Attractor (green double-circled nodes) and repeller strings (red nodes).}
    \label{fig:quine}
\end{figure}

In our model, we consider the entire set of $2^l$ strings.
Each string is represented by a node in Figure~\ref{fig:quine}, with arrows indicating the mapping: running the string as a program on the automata.
Thus, while many-to-one arrows are possible, one-to-many is not.
We partition the set of strings (interpreted as program or data) into 2 subsets: attractors and repellers.
Attractors are strings that, when executed as a program, generate as output a string that is also from the attractor subset.
A repeller string, when run as a program, can generate either an attractor or a repeller string.
The entire space might have multiple connected components.
Each connected component consists of an attractor basin, made of quines or quine-relays (as the multi-node attractor cycles) in Figure~\ref{fig:quine}.
Each node in the attractor basin might have a trail of repeller nodes that, over cycles of algorithmic probability, converge to the basin node.

Over multiple cycles, the final set will include only the attractor nodes, with a one-to-one mapping, thus conserving the number of strings in subsequent cycles.
We denote this specific number of cycles with $w_t$, the meta-level at which a uniform distribution results in only attractors after $w_t$ cycles.
Each repeller node can be assigned a number equal to the number of hops from an attractor basin.
The highest hops are the $w_t$, i.e. $6$ in the example shown in the figure.
At this stage, each string has a one-to-one mapping.
The number of attractors is given by 
$$Q = N^{0w_t} = |\xi^{0w_t}(x^0) > 0|$$
The algorithmic probability for these constructor strings depends on the number of paths leading to these attractor basins over these cycles.
$w_t$ is dependent on the number of considered states and symbols, the specification of the automata, the length of the programs, and the time approximation for estimating the algorithmic probability at each level.
Being at least as (semi-) uncomputable as $\xi$, we can only study it under reasonable approximations of experimental algorithmic information theory (EAIT)~\cite{Delahaye2007OnTK,zenil_2020}.

\subsection{Empirical experiments on linear-bounded automata}

Here, we consider a specific case to illustrate the formalism for nested algorithmic probability.
We take the $ 2$-state $ 2$-symbol linear-bounded automata~(LBA) as it is both non-trivial and within the bounds of exhaustive enumeration.
The details of this machine can be found in \cite{sarkar2020quantum}.
The program (description number) is encoded as the list of transition functions for each state and read symbol:
$$[QMW]^{Q_1R_1}[QMW]^{Q_1R_0}[QMW]^{Q_0R_1}[QMW]^{Q_0R_0}$$
This gives $q_\delta = 12$, the length of the description number required to store a program for this machine.
Thus, the space of this encoding allows $P = 2^{12} = 4096$ possible programs.
The tape is also of length $c = z = 12$ and consists of all zeros with the tape head on the left-most character: $\underline{o}ooooooooooo$
The machine is run for $z = q_\delta = 12$ iterations.
A Python script that emulates this automata model for all $4096$ cases is available at \cite{QuBio_2020}.

\begin{figure}[htb]
\centering
\includegraphics[clip, trim=0cm 0cm 0cm 0cm, width=0.6\columnwidth]{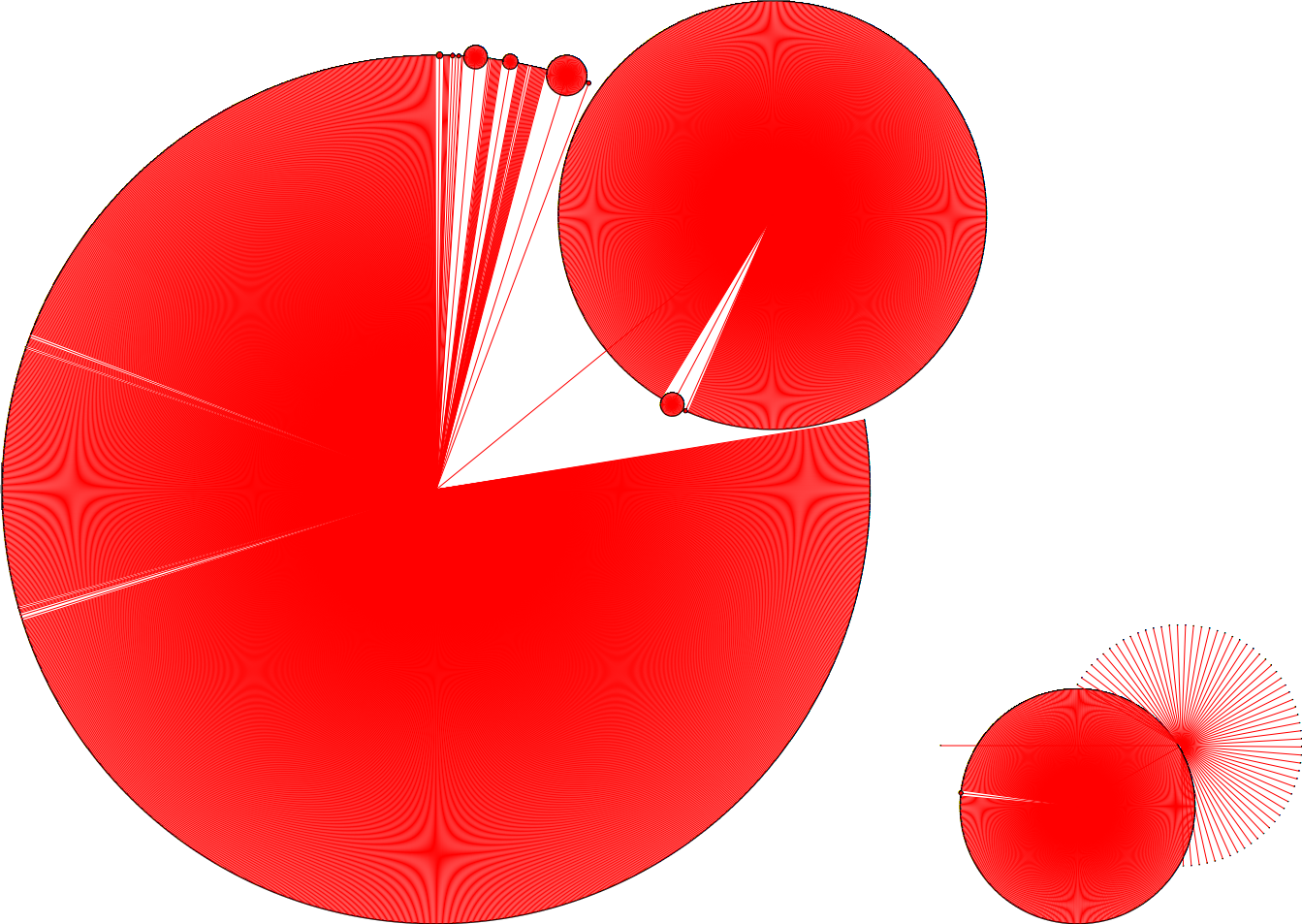}
\caption{Level 3 of nesting algorithmic probability for a 2-state 2-symbol LBA (the high-resolution SVG is available at \cite{QuBio_2020}).}
\label{fig:map3}
\end{figure}

At level 3 of the nesting algorithmic probability, we start with a uniform distribution of all programs to produce the standard universal distribution.
The mapping is shown in Figure~\ref{fig:map3} (the high-resolution SVG is available at \cite{QuBio_2020}).

\begin{figure}[htb]
\centering
\includegraphics[clip, trim=7.4cm 0cm 0cm 0cm, width=0.6\columnwidth]{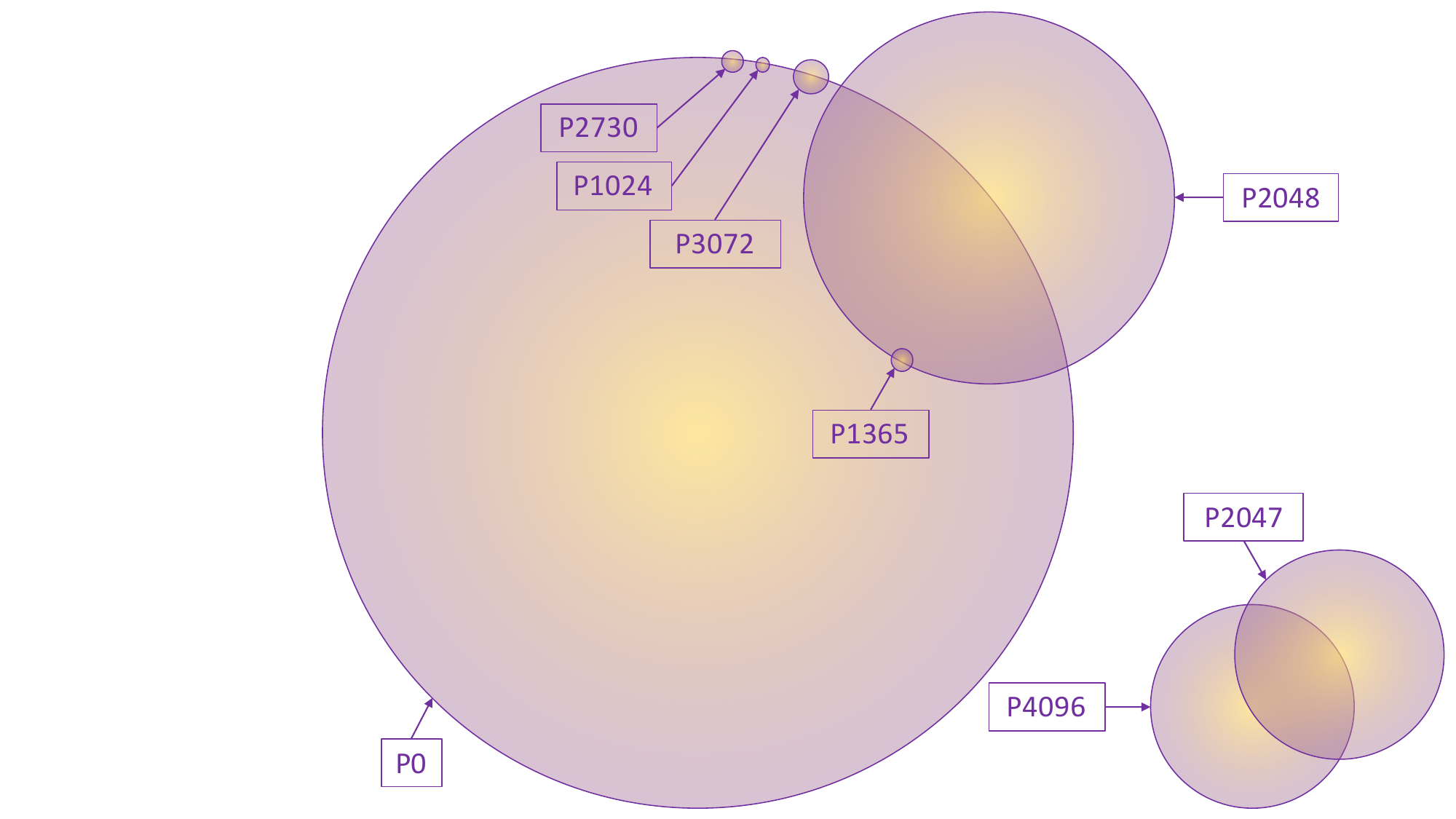}
\caption{The 8 largest attractor basins of level 3 of the nesting algorithmic probability.}
\label{fig:map3basin}
\end{figure}

We observe that, at this level itself, the number of possible programs for the next generation gets reduced from $4096$ to $21$, with the following frequency of occurrence.
The top 8 algorithmically probable attractor basins are shown in Figure~\ref{fig:map3basin}.
\begin{verbatim}
   P0    : 1886     P2048 : 1147     P4095 : 640      P3072 : 110
   P1365 : 64       P2047 : 64       P2730 : 64       P1024 : 41
   P3840 : 17       P128  : 11       P3968 : 11       P1344 : 10
   P3584 : 10       P4032 : 10       P1792 : 2        P1920 : 2
   P2560 : 2        P2688 : 2        P192  : 1        P1728 : 1
   P2944 : 1
\end{verbatim}
Also, at this level itself, we find that the machines P0 and P4095 are the only self-replicating programs, thus the limiting behavior can already be predicted.

At level 2, these $21$ programs get further mapped to just $3$ programs, with the following frequency.
\begin{verbatim}
   P0    : 16       P4095 : 3        P2048 : 2
\end{verbatim}

\begin{figure}[hbt]
\centering
\includegraphics[clip, trim=0cm 0cm 0cm 0cm, width=0.6\columnwidth]{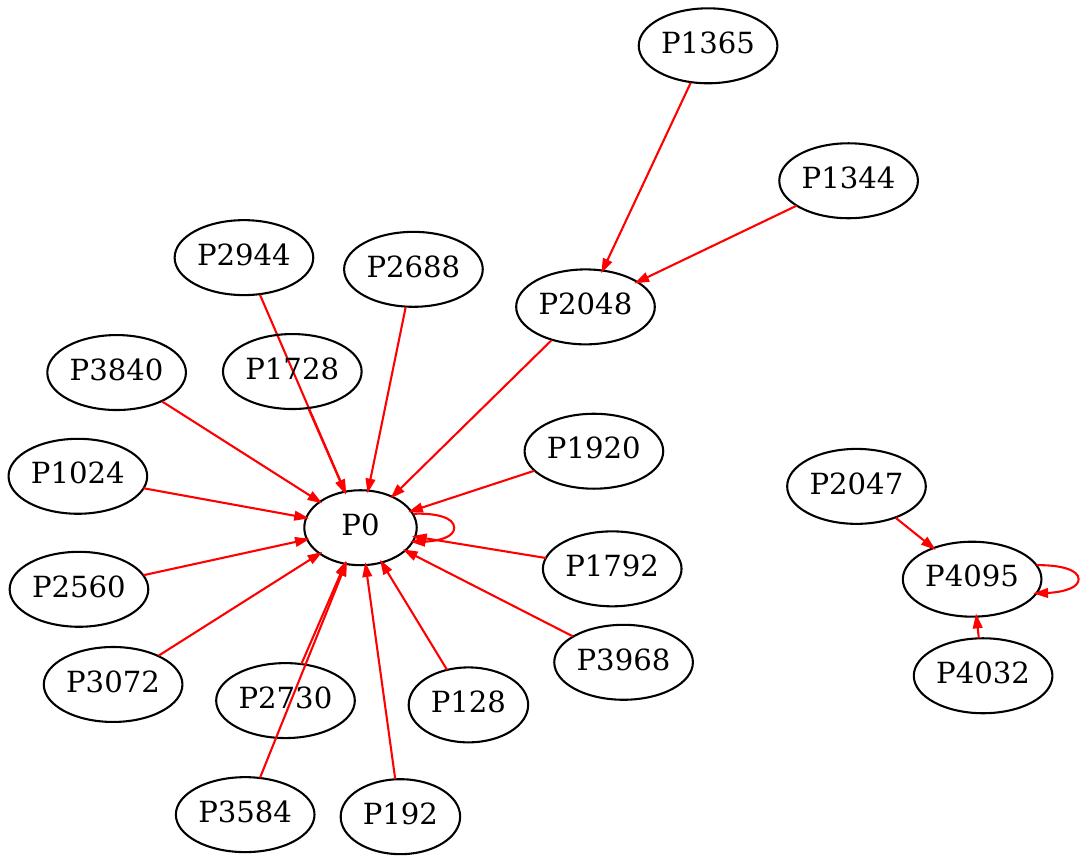}
\caption{Level 2 of nesting algorithmic probability for a 2-state 2-symbol LBA.}
\label{fig:map2}
\includegraphics[clip, trim=0cm 0cm 0cm 0cm, width=0.25\columnwidth]{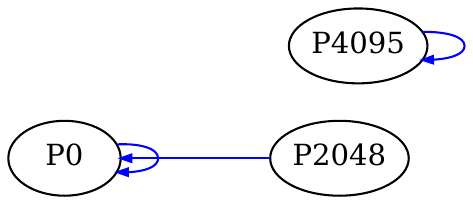}
\caption{Level 1 of nesting algorithmic probability for a 2-state 2-symbol LBA.}
\label{fig:map1}
\includegraphics[clip, trim=0cm 0cm 0cm 0cm, width=0.18\columnwidth]{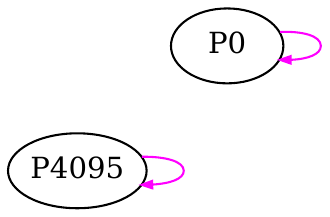}
\caption{Level 0 of nesting algorithmic probability for a 2-state 2-symbol LBA.}
\label{fig:map0}
\end{figure}

At level 1, the $3$ programs finally converge to the quines P0 and P4095.


The frequency is as follows:
\begin{verbatim}
   P0    : 2        P4095 : 1
\end{verbatim}

At level 0, we reach the attractor states with a uniform one-to-one mapping.

Level 2, 1 and 0 are shown in Figure~\ref{fig:map2},\ref{fig:map1} and \ref{fig:map0} respectively.


Now, to calculate the overall algorithmic probability of these two fixed points, we compute the cumulative frequency across the 4 levels.
Thus, at level 2, the total frequency of P4095 consists of adding up the frequency of programs P4095, P2047, P4032 from the previous step, totaling to $640+64+10 = 714$.
For P2048, we add the frequencies of Programs P1365 and P1344, totaling $ 64 + 10 = $74.
The remaining $16$ programs total $3308$ cases that reach P0.
At level 1, P2048 also reaches the P0 attractor, giving the final algorithmic frequency of the quines as:
\begin{verbatim}
   P0    : 3382     P4095 : 714
\end{verbatim}
We call this the resource-bounded Universal Constructor distribution.

A similar experimentation on this space of 4096 programs is conducted as part of the Wolfram Physics Project exploring the rulial space of Turing machines~\cite{wolfram_2020}.
The 2-2-1 LBA with 4096 programs shows little diversity, resulting in simple quines.
It remains to be seen what the encodings of quines in larger spaces reveal about the structure and complexity of constructors.

\section{Discussions}
\label{sec_4}


This work provides a formal explanation for the spontaneous emergence and persistence of self-replication in abstract computational systems. 
By introducing resource-bounded nested algorithmic probability, we showed that universal induction under finite constraints induces an effective ecological process over programs. 
Within this process, self-constructing programs, quines and quine-relays, emerge as stable fixed points and dominate the resulting distribution. 
Crucially, this dominance does not arise from optimization, selection, or externally imposed objectives and rewards, but from the intrinsic structure of universal computation under bounded resources. 

Our work serves as a foundation that unifies and explains recent empirical observations in computational life and artificial intelligence, showing that self-replicating programs emerge ubiquitously in minimal computational substrates.
We interpret these results~\cite{alakuijala2024computational,chaitin2013proving,sharma2023assembly,zhang2025darwin,kumar2026digital} as empirical confirmations of the theoretical principle developed here: once a system supports universal computation and is subject to resource bounds, self-replication is not merely possible but statistically inevitable.

The implications of this formalism extend across multiple domains. 
In biology, particularly synthetic and xenobiology~\cite{sarkar2021estimating,marletto2015constructor,sipper2001go}, the results suggest that replication is not a fragile chemical coincidence but a generic consequence of constrained information-processing systems. 
For artificial general intelligence, including large language models and universal learning agents, the framework can be extended to reason about why self-modeling~\cite{premakumar2024unexpected}, tool construction~\cite{wang2023voyager}, and recursive improvement emerge from a computational resource perspective. 
In physics, the results resonate with constructor theory~\cite{deutsch2013constructor} by identifying self-replication as the simplest stable structures permitted by the laws of computation under constraint, suggesting a computational underpinning~\cite{toffoli2003lagrangian,wolfram2022physicalization} for physical persistence and causation.

Taken together, our results support a strong conceptual claim: life is the simplest persistent structure in bounded computational universes. 
A system that constructs and maintains a description of itself necessarily minimizes its effective entropy production relative to its environment. 
In this sense, remembering oneself is not an adaptive strategy layered atop physics, but a mathematical identity arising from constrained universal computation. 
To summarize:

\begin{center}
\textit{\large``Life is not what computation does when optimized;\\it is what computation does when constrained."}  
\end{center}
\vspace{1em}